\documentstyle[aps,prl,twocolumn,floats,epsf,amsfonts]{revtex}
\newcommand{\ph}{{\phantom{\dagger}}}
\begin{document}
\draft
\title{Flow equation solution for the weak to strong--coupling\\ crossover in the sine--Gordon model}
\author{Stefan Kehrein \cite{leave}}
\address{Lyman Laboratory of Physics, Harvard University, Cambridge, MA 02138\\
{\rm(October 12, 1999)}
}

\address{~
\parbox{15cm}{\rm
\medskip
A continuous sequence of infinitesimal unitary transformations, combined with an operator
product expansion for vertex operators, is used to diagonalize the quantum sine--Gordon
model for $\beta^2\in(2\pi,\infty)$. The leading order of this approximation already 
gives very accurate results for
the single--particle gap in the strong--coupling phase. This approach can be understood
as an extension of perturbative scaling theory since it links weak to strong--coupling
behavior in a systematic expansion.
The method should also be useful for other strong--coupling problems that can be
formulated in terms of vertex operators. 
\\~\\
PACS numbers: 71.10.Pm, 11.10.Gh, 11.10.Hi
}}

\maketitle

\narrowtext
Perturbative scaling arguments play an important role for analyzing a large variety of 
physical systems with many degrees of freedom. For strong--coupling problems, however, 
the perturbative renormalization group (RG) equations lead to divergences in the running coupling 
constants and the perturbative RG--approach becomes invalid. In condensed matter 
theory the well--known paradigm for this kind of behavior is the Kondo model: 
The perturbative scaling equations still allow one to identify the low--energy scale of the 
Kondo model, but by themselves they do not lead to an understanding of the physical 
behavior associated with this energy scale. Wilson's numerical RG \cite{kondo}     
could remedy this problem, but an analytical RG--like approach that links weak to
strong--coupling behavior in an expansion that can be systematically improved would still 
be desirable for many strong--coupling problems.

In this Letter it will be shown exemplary how Wegner's {\em flow equations} 
\cite{Wegner94} can provide such an analytical description for a weak to strong--coupling 
behavior crossover. In the flow equation approach a continuous
sequence of infinitesimal unitary transformations is employed to make a 
Hamiltonian successively more diagonal. Large energy differences are decoupled
before smaller energy differences, which makes the method similar to the
conventional RG~approach. However, degrees of freedom are not integrated out
as in the RG but instead diagonalized. A similar framework that contains
Wegner's flow equations as a special case has independently been developed
by G{\l}azek and Wilson ({\em similarity renormalization scheme}) 
\cite{GlazekWilson}.

The model under investigation in this Letter is
the $1+1d$~quantum sine--Gordon model~\cite{rev_sinegordon}
\begin{displaymath}
H=\int dx\, \left( \frac{1}{2}\Pi(x)^2+\frac{1}{2}\left(\frac{\partial\phi}{\partial x}\right)^2
+u \tau^{-2} \cos\left[\beta\phi(x)\right] \right)
\end{displaymath}
with the commutator $[\Pi(x),\phi(x')]=-i\delta(x-x')$. Regularization with
a UV--momentum cutoff~$\Lambda\propto\tau^{-1}$ is implied and $u,\beta>0$ 
are dimensionless parameters.
The sine--Gordon model is one of the best studied integrable models and
it has been solved using the inverse scattering method~\cite{Faddeev75}. 
This model is therefore a good test case for studying the new approach.
We will be interested in the universal low--energy properties ($E\ll\Lambda$)
for small coupling constants~$u$. 
It should be emphasized that the integrable structure underlying the inverse scattering
solution will {\em not} be used in the approximate flow equation solution; the new
method can also be used when non--integrable perturbations are added.

The phase diagram of the sine--Gordon model consists of a  
gapped phase for $\beta^2\lesssim 8\pi$ with massive soliton excitations
and a gapless phase for $\beta^2\gtrsim 8\pi$ with massless solitons \cite{rev_sinegordon}.
The phase transition between these two phases for $\beta^2/8\pi= 1+O(u)$ is of the
Kosterlitz--Thouless type. In the massive phase the perturbative scaling equations 
\cite{Wiegmann} lead to an unphysical strong--coupling divergence of the running coupling constant~$u$.
The inverse scattering solution \cite{Faddeev75} furthermore shows the existence of bound
soliton states (breathers) in the spectrum for $\beta^2<4\pi$ while such bound states are absent
for $\beta^2>4\pi$. For $\beta^2=4\pi$ the sine--Gordon model can be mapped to a 
noninteracting massive Thirring model \cite{Coleman}, which in turn can be
diagonalized easily leading to the identification of the massive solitons with
the Thirring fermions \cite{Mandelstam}. The sine--Gordon model is also related
to a variety of other models like the spin-$1/2$ X-Y-Z chain, a $1d$~Fermi system with
backward scattering 
and the $2d$~Coulomb gas with temperature $T=\beta^{-2}$ and fugacity $z\propto u$,
the IR--unstable fixed point corresponds to $T=1/8\pi$ \cite{Solyom}.

It will be shown that the flow equation approach generates a diagonalization of the sine--Gordon 
Hamiltonian both in the weak--coupling and in the strong--coupling phase
in a systematic expansion that can be successively improved: No divergences of the
running couplings are encountered in the strong--coupling regime for $\beta^2>2\pi$.
The soliton mass is found to be in very good agreement with the inverse scattering solution.
The crossover from weak-- to strong--coupling behavior can be described and using this
the soliton dispersion relation for example can be analyzed on all energy scales. 

For the purposes of this Letter it will be more convenient not to use a regularization of 
the sine--Gordon model with an explicit momentum cutoff, but instead to ``smear out'' 
the interaction term
\begin{eqnarray}
H&=&\int dx\, \Bigg( \frac{1}{2}\Pi(x)^2+\frac{1}{2}\left(\frac{\partial\phi}{\partial x}\right)^2 
\label{sineGordon} \\
&&\qquad +\frac{u}{\pi a^2} \cos\left[\beta \int dy\, c(y)\,\phi(x+y) \right] \Bigg) \nonumber
\end{eqnarray}
with the Lorentzian
$c(y)=a/(2\pi y^2+\pi a^2/2)\: \stackrel{a\rightarrow 0}{\longrightarrow}\; \delta(y)$
\cite{regularization}.
This does not affect the universal properties for small energies $E\ll\Lambda\propto a^{-1}$.
We expand the fields in normal modes
\begin{eqnarray}
\phi(x)&=&-\frac{i}{\sqrt{4\pi}} {\sum_{k\neq 0}} \,\frac{\sqrt{|k|}}{k}\, e^{-ikx}
\left(\sigma_1(k)+\sigma_2(k)\right) \nonumber \\
\Pi(x)&=&\frac{1}{\sqrt{4\pi}}{\sum_{k\neq 0}}\,\sqrt{|k|}\, e^{-ikx}
\left(\sigma_1(k)-\sigma_2(k)\right) \nonumber
\end{eqnarray}
where ${\sum_k}\stackrel{\rm def}{=}\frac{2\pi}{L}\sum_{n=-\infty}^\infty$ with
$k=2\pi n/L$. $L$ is the system size. The basic commutators are ($k,k'>0$)
$[\sigma_1(-k),\sigma_1(k')]=[\sigma_2(k),\sigma_2(-k')]=\delta_{kk'}\, L/2\pi$,
notice $\sigma_i^\dagger(-k)=\sigma_i^{\phantom\dagger}(k)$. The vacuum is defined 
by $\sigma_1(-k)|\Omega\rangle=\sigma_2(k)|\Omega\rangle=0$ for $k>0$.

The flow equation approach \cite{Wegner94} (see also \cite{rev_floweq})
generates a family of unitarily equivalent Hamiltonians $H(B)$ as a function of 
a flow parameter~$B$ (with dimension (Energy)${}^{-2}$), where $H(B=0)$ is the initial 
Hamiltonian and $H(B=\infty)$ the final diagonal Hamiltonian. 
This flow is generated by the differential equation
\begin{equation}
\frac{dH(B)}{dB}=[\eta(B),H(B)]
\label{flowequation}
\end{equation}
with $\eta(B)=-\eta(B)^\dagger$ some antihermitean generator. Generically,
Eq.~(\ref{flowequation}) leads to the generation of new interaction terms not contained 
in the initial Hamiltonian. We therefore write 
$H(B)=H_0+H_{\rm int}(B)+H_{\rm new}(B)$ with
\begin{eqnarray}
H_0&=&{\sum_{p>0}} \, p\, \big(\sigma_1(p)\sigma_1(-p)+\sigma_2(-p)\sigma_2(p)\big) \nonumber \\
H_{\rm int}(B)&=&\int dx_1\,dx_2\:u(B;x_1-x_2)  
\label{Hint} \\
&&\quad\times\big(V_1(\alpha;x_1)V_2(-\alpha;x_2)+{\rm h.c.}\big) \ .
\nonumber 
\end{eqnarray}
Here $V_j(\alpha;x)$ are normal ordered vertex operators
with scaling dimension $\alpha(B)\stackrel{\rm def}{=}\beta(B)/\sqrt{4\pi}$
\begin{displaymath}
V_j(\alpha;x)=\; :\exp\Big(\pm\alpha{\sum_{p\neq 0}}\frac{\sqrt{|p|}}{p}e^{-\frac{a}{2}|p|-ipx}
\sigma_j(p)\Big):
\end{displaymath}
where $+$ (upper sign) for $j=1$ and $-$ (lower sign) for $j=2$.
To avoid confusion the initial values of the couplings will from now on be denoted by $u_0, \beta_0$.
$H(B=0)$ is identical to the sine--Gordon Hamiltonian~(\ref{sineGordon}) for 
$H_{\rm new}(B=0)=0$, $\alpha(B=0)=\beta_0/\sqrt{4\pi}$ and 
$u(B=0;x)=u_0\,\delta(x)\, (2\pi a/L)^{\alpha(B=0)^2}/2\pi a^2$.

Wegner's idea for constructing a suitable generator $\eta$ is to choose
$\eta(B)\stackrel{\rm def}{=}[H_0,H_{\rm int}(B)]$ \cite{Wegner94}. This gives
\begin{displaymath}
\eta=-2i\int dx\,dy\,(\partial_y u(y))
\big(V_1(\alpha;x)V_2(-\alpha;x-y)+{\rm h.c.}\big) .
\end{displaymath}
Using this generator, matrix elements connecting states with large energy differences
are eliminated for small~$B$, while matrix elements coupling more degenerate states 
are eliminated later for larger values of~$B$. 

First we evaluate $[\eta,H_0]$. This leads to
$\partial u/\partial B=4\,\partial^2 u/\partial x^2$,
which makes the interaction increasingly non--local along the
flow due to the decoupling procedure. In 
Fourier components $u(B;x)={\sum_p}\, u_p(B)e^{-ipx}$ one finds
\begin{displaymath}
u_p(B)=\frac{\tilde u(B)}{4\pi^2 a^2}\,
e^{-4p^2B} \left(\frac{2\pi a}{L}\right)^{\alpha^2} \ .
\end{displaymath}
$\tilde u(B)$ will turn out to be the running coupling constant of the flow equation approach
and remains finite also in the strong--coupling phase. 
$\tilde u(B)$ is dimensionless, initially $\tilde u(B=0)=u_0$. The term $[\eta,H_{\rm int}]$ 
leads to new interactions that have to
be truncated to obtain a closed set of equations. The approximation used here is to take
only operators with small scaling dimensions into account, i.e. to neglect more irrelevant
operators. It is generated by truncating the operator product expansion (OPE) of two
vertex operators in the following way
\begin{eqnarray}
\lefteqn{ V_j(\alpha;x)V_j(-\alpha;y)=\left(\frac{L}{2\pi}\right)^{\alpha^2}
\frac{1}{\left[a\mp i(x-y)\right]^{\alpha^2}} }
\label{OPE} \\
&\times&\Big(1\mp i\alpha(x-y) {\sum_{p\neq 0}} \sqrt{|p|} \, e^{-\frac{a}{2}|p|-ipx}
\sigma_j(p) +\ldots\Big) \ . \nonumber
\end{eqnarray}
The approximation can be systematically improved by going to higher orders in this OPE. 
The term $[\eta,H_{\rm int}]$ contains commutators with the structure 
\begin{eqnarray}
\lefteqn{[V_1(\alpha;z_1)V_2(-\alpha;z_2),V_2(\alpha;z_2')V_1(-\alpha;z_1')]}
\label{etaHint} \\
&=&-\{V_1(\alpha;z_1),V_1(-\alpha;z_1')\}V_2(\alpha;z_2')V_2(-\alpha;z_2) \nonumber \\
&&+V_1(\alpha;z_1)V_1(-\alpha;z_1')\{V_2(\alpha;z_2'),V_2(-\alpha;z_2)\} \nonumber
\end{eqnarray}
and terms where $\alpha\rightarrow -\alpha$ in one argument of 
the commutator~(\ref{etaHint}). After normal ordering, the latter terms lead to
interactions $V_1(2\alpha;z_1)V_2(-2\alpha;z_2)$ with larger scaling dimensions.
These will be neglected, but the terms generated by~(\ref{etaHint}) will be included.

For $\alpha=1$ ($\beta^2=4\pi$) the vertex
operators describe fermions and the OPE~(\ref{OPE}) to all orders gives
$\{V_j(-1;x),V_j(1;y)\}\stackrel{a\rightarrow 0}{\longrightarrow} L\delta(x-y)$.
Since this is a $c$--number, no higher order interactions are generated in
(\ref{etaHint}) and the flow equations {\em close}. The flow equations therefore 
recover the equivalence of a sine--Gordon model with $\beta_0^2=4\pi$ 
to a massive noninteracting Thirring model~\cite{Coleman} and readily diagonalize the
latter.

In general we evaluate (\ref{etaHint}) using (\ref{OPE}). The dominating contributions 
decaying most slowly with~$B$ can be identified in closed form \cite{details}. 
Two structurally different interaction terms are generated: One term contributes to $H_{\rm new}$
and is discussed below (see Eq.~(\ref{Hdiag})). The other term has the structure 
$\sigma_1(k)\sigma_2(-k)$. Integrating it from $B$ to $B+dB$,
one generates ${\sum_k} w_k |k| \sigma_1(k)\sigma_2(-k)$ 
with infinitesimal coefficients~$w_k$. This new interaction can be removed by
a further infinitesimal unitary transformation with the structure $e^{-U} H(B) e^{U}$, 
where $U=\frac{1}{2}{\sum_{k>0}} w_k ( \sigma_1(k)\sigma_2(-k)-{\rm h.c.})$.
This new infinitesimal unitary transformation yields a modification of the 
scaling dimension of the vertex operators in $H_{\rm int}$ and a flow of the
coupling constant~$\tilde u(B)$. In terms of a logarithmic dimensionless flow parameter 
$\ell=\frac{1}{2}\ln\big(32B a^{-2}\big)$ one derives \cite{details}
\begin{eqnarray}
\frac{d\beta^2(\ell)}{d\ell}&=&-u_0^2\, 
\frac{\beta^4(\ell)}{4\pi\,\Gamma\big(-1+\beta^2(\ell)/4\pi\big)}
\label{flowbeta2} \\
&&\quad\times\exp\left(4\ell-\frac{1}{2\pi}\int_0^\ell d\ell'\, \beta^2(\ell') \right) \nonumber \ .
\end{eqnarray}
The running coupling constant flows according to
$\tilde u(\ell)=u_0\, \exp\big(F(\ell)\big)$, where
\begin{equation}
F(\ell)=\frac{1}{4\pi}\left( \ell\,\beta^2(\ell)
-\int_0^\ell d\ell'\,\beta^2(\ell') \right) \ .
\label{eq_F}
\end{equation}
In the strong--coupling phase the flow terminates at $\beta^2(\infty)=4\pi$
due to the divergent $\Gamma$--function in (\ref{flowbeta2}). In 
our approach $\beta^2=4\pi$ is therefore an {\em attractive} strong--coupling
fixed point. 
For comparison with the RG--equations \cite{Wiegmann} one can introduce 
$u(\ell)\stackrel{\rm def}{=} u_0\, 
\exp\left(2\ell-\frac{1}{4\pi}\int_0^\ell d\ell'\,\beta^2(\ell')\right)$
and rewrite (\ref{flowbeta2}) as two coupled differential equations
\begin{eqnarray}
\frac{d\beta^{-2}(\ell)}{d\ell}&=&\frac{1}{4\pi\,
\Gamma\big(-1+\beta^2(\ell)/4\pi\big)}\,u^2(\ell)
\label{compRG} \\
\frac{du(\ell)}{d\ell}&=&\left(2-\frac{\beta^2(\ell)}{4\pi}\right) u(\ell) \nonumber
\end{eqnarray}
with $\beta(\ell=0)=\beta_0, u(\ell=0)=u_0$.
For $\beta_0^2=8\pi$ Eqs.~(\ref{compRG}) coincide with the two loop scaling equations 
\cite{Wiegmann}:
Depending on the value of $u_0$, the sine--Gordon model for $\beta_0^2>8\pi$
flows to either $\beta^2(\infty)=4\pi$ ({\em strong--coupling}) or $\beta^2(\infty)\geq 8\pi$
({\em weak--coupling}). Eqs.~(\ref{compRG}) therefore reproduce the Kosterlitz--Thouless 
phase diagram. Also the hidden SU(2)--symmetry in (\ref{sineGordon}) for 
$\beta_0^2= 8\pi(1\pm u_0), u_0\ll 1$
is recovered although our approximation scheme does not manifestly respect this symmetry \cite{SU2}.

In the strong--coupling phase a gap opens in the spectrum and the low--energy excitations
are fermionic \cite{Faddeev75}. In our approach this follows most easily by approximating
(\ref{Hint}) for large~$B$ (such that $|\beta^2(B)-4\pi|\ll 1$) as
\begin{eqnarray}
H_{\rm int}(B)&=&\int dx_1\,dx_2\, \frac{\tilde u(B)}{\pi a^2}
\:\frac{1}{\sqrt{16\pi B}}\exp\left(-\frac{(x_1-x_2)^2}{16B}\right) \nonumber \\ 
&\times&\left( \psi_1^\dagger(x_1) \psi_2^\ph (x_2) + \psi_2^\dagger(x_2) \psi_1^\ph (x_1) \right)
\label{Hint2}
\end{eqnarray}
with fermions $\psi_j(x)\stackrel{\rm def}{=}V_j(-1;x)$.
The running coupling constant $\tilde u(B)$ approaches a 
{\em finite} value in this limit \cite{divv}. The asymptotic value $\tilde u(\infty)$
can therefore be interpreted as the renormalized coupling constant parametrizing a
quadratic Hamiltonian $H_0+H_{\rm int}(B)$ that describes the low--energy behavior of the
initial sine--Gordon model. The RG strong--coupling divergence of the running coupling
constant is avoided since the interaction (\ref{Hint2}) in this effective low--energy Hamiltonian 
becomes increasingly non--local for $B\rightarrow\infty$. The gap 
$\Delta=2m$ in the spectrum is obtained in terms of the renormalized coupling constant 
$\tilde u(\infty)$ that sets the mass of the effective low--energy Thirring fermions with
$m=\tilde u(\infty)/a$.

\begin{figure}[t]
\begin{center}
\leavevmode
\epsfxsize=7cm
\epsfysize=6cm
\epsfbox{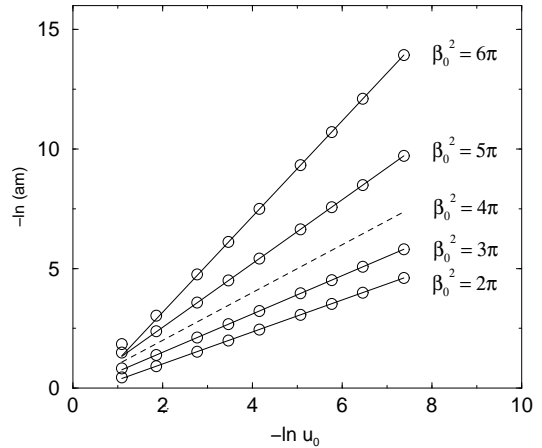}
\caption{Soliton mass as a function of the coupling constant for various
values of~$\beta_0^2$: The full lines are constrained fits of the power
law behavior $am\propto u_0^{1/(2-\beta_0^2/4\pi)}$~\protect\cite{rev_sinegordon} 
to the flow equation
results (open circles) with the proportionality constant being fitted. The
dashed line is the case $\beta_0^2=4\pi$ where the flow equation
approach agrees trivially (see text).}
\end{center}
\vspace*{-0.7cm}
\end{figure}

The soliton mass derived from Eqs.~(\ref{eq_F}-\ref{compRG}) can be compared
with the predictions of perturbative scaling:
E.g.\ for $\beta_0^2=8\pi(1-u_0)$ (corresponding to the hidden SU(2)--symmetry)
one can show analytically $m\propto u_0^\tau \exp(-1/2u_0) /a$ for $u_0\downarrow 0$ 
with $\tau=(1+\gamma)/3\approx 0.526$ ($\gamma\approx 0.577$ is Euler's constant). 
This agrees well with the third order RG result
$\tau_{\rm RG}=1/2$ and the same exponential dependence \cite{Wiegmann,alpha}.
For smaller values of~$\beta_0$ one finds the
power law behavior $m\propto u_0^{1/(2-\beta_0^2/4\pi)}/a$ known from the 
inverse scattering solution \cite{rev_sinegordon}: This is very well confirmed by the 
numerical integration of the flow equations (Fig.~1). The 
proportionality constant depends only on~$\beta_0$ for $u_0\downarrow 0$,
in particular $m=u_0/a$ for $\beta_0^2=4\pi$ as known
from the noninteracting Thirring model \cite{Coleman}. 

The flow equation solution not only provides the excitation gap, but one can also obtain information 
about the crossover, e.g.\ the full dispersion relation:  
The Hamiltonian for $B=\infty$ takes the form $H(\infty)=H_0+H_{\rm new}(\infty)$ 
since the interaction term $H_{\rm int}(B)$ is eliminated. One can verify
that $H_{\rm new}(B)$ does not modify the flow of $\beta(B)$ and $\tilde u(B)$ as derived above \cite{details}. 
The new terms generated during the flow in $H_{\rm new}(\infty)$ can be split up as
$H_{\rm new}(\infty)=H_{\rm diag}(\infty)+H_{\rm res}(\infty)$, where
$H_{\rm diag}(\infty)$ contains the terms that follow in leading order of the
OPE from $[\eta,H_{\rm int}]$ in~(\ref{etaHint}), while $H_{\rm res}(\infty)$ formally
contains everything not taken into account in the present order of the OPE. 
Integration of the flow equations gives~\cite{details}
\begin{eqnarray}
H_{\rm diag}(\infty)&=&\sum_{p>0} \omega_p(\infty) \Big( P_1^\dagger(p) P_1^{\phantom\dagger}(p)
+ P_1^{\phantom\dagger}(-p) P_1^{\dagger}(-p)
\nonumber \\
&&\qquad + P_2^\dagger(-p) P_2^{\phantom\dagger}(-p)
+ P_2^{\phantom\dagger}(p) P_2^{\dagger}(p) \Big) 
\label{Hdiag}
\end{eqnarray}
with certain coefficients $\omega_p(\infty)$. Here
$P_j(p)=\int dx\:e^{-ipx} V_j(-\alpha(B_p);x)$
and $B_p\stackrel{\rm def}{=}1/4p^2$. The spectrum can be analyzed easily
since $[H_0,H_{\rm diag}(\infty)]=0$: In leading order the single--particle 
(soliton) excitations of $H_0+H_{\rm diag}(\infty)$ are $P_1^\dagger(k)|\Omega\rangle$ 
for $k>0$ and $P_2^\dagger(k)|\Omega\rangle$ for $k<0$. The single--hole (antisoliton) excitations are 
$P_1(k)|\Omega\rangle$ for $k<0$ and $P_2(k)|\Omega\rangle$ for $k>0$. In the strong--coupling phase
for $\beta_0^2\geq 4\pi$ the resp.\ excitation energies are very accurately (but not exactly) described by 
$E_k^2=k^2+(\tilde u(\infty)/a)^2$ in the small coupling limit: 
There are $\beta_0$--dependent universal corrections in the crossover region $k=O(m)$ 
that vanish for $\beta_0^2\rightarrow 4\pi$ and reach at most of order $2\%$ 
for $\beta_0^2=8\pi$.
The character of the excitations varies
from scaling dimension~$\alpha(B=0)$ to the low--energy Thirring fermions with $\alpha(B=\infty)=1$.  
In the weak--coupling phase the spectrum remains gapless and $E_k=|k|$ for $k\rightarrow 0$.
Notice that the elementary excitations are expressed with respect to a transformed basis 
since $H(\infty)$ and $H(0)$ are related by a complicated unitary transformation.

For $\beta_0^2<2\pi$ the differential equations for $\omega_p(B)$ lead to 
divergences since then the $\cos(\beta_0\phi(x))$--perturbation
is too relevant, limiting the approach to $\beta_0^2>2\pi$. For
$2\pi<\beta_0^2<4\pi$ the single--particle spectrum is still well described by (\ref{Hdiag}),
our approximations become better as $\beta_0^2\uparrow 4\pi$. Higher orders in the OPE are
nevertheless required to study the formation of bound states 
for $\beta_0^2<4\pi$ \cite{Faddeev75} due to residual interactions in $H_{\rm res}(\infty)$.
 
Summing up, we have applied a continuous sequence of infinitesimal unitary transformations,
combined with an operator product expansion for vertex operators, to the quantum sine--Gordon
model with $\beta^2\in(2\pi,\infty)$. The approximations are systematic since
more terms in the OPE can successively be taken into account and will not endanger the 
stability of the strong--coupling fixed point. The results for the
soliton mass in the strong--coupling phase agree with 
two loop scaling predictions for $\beta^2\approx 8\pi$ (approximate agreement was even found 
to three loop order) and exact methods \cite{rev_sinegordon} applicable 
for smaller~$\beta^2$. The full dispersion relation could be obtained and the crossover from
weak to strong--coupling behavior described. The method 
also allows one to study correlation functions \cite{KehreinMielke} and
other strong--coupling problems that can be formulated in terms of vertex operators.

The author acknowledges many valuable discussions with K.~Byczuk, T.~Franosch, 
W.~Hofstetter and particularly D.~S.~Fisher.
This work was supported by the Deutsche Forschungsgemeinschaft (DFG) and
the National Science Foundation (NSF) under grants DMR~9630064 and DMR~9976621.
\vspace*{-0.5cm}


\begin{references}
\bibitem{details} Details of this calculation will be presented elsewhere.
\bibitem{SU2} This hidden symmetry has its origin in the SU(2)--spin symmetry of 
a $1d$~Fermi system with backward scattering that can be mapped to the sine--Gordon model  
(\ref{sineGordon}) with $\beta_0^2=8\pi(1\pm u_0)$, $u_0\ll 1$ \protect\cite{Wiegmann,Solyom}.
\bibitem{alpha} Higher orders in the OPE can modify~$\tau$, 
therefore the already  
small difference to~$\tau_{\rm RG}$ is remarkable.
\end{references}
\end{document}